# CEPC Project Status

*Jie* Gao[1,2,*]
[1]Institute of High Energy Physics, 100049 Beijing, China
[2]University of Chinese Academy of Sciences, 100049 Beijing, China

**Abstract.** In this article we give a brief historical review of particle physics achievement through e+e- colliders and proton accelerators in the last half century and perspective for the next half century with worldwide consensus. The Circular Electron Positron Collider (CEPC) is a Higgs factory proposed by Chinese scientists in 2012, and its Technical Design Report (TDR) has been completed in Dec. 2023. Since 2024 CEPC has entered into Engineering Design Report (EDR) phase (2024-2027) and CEPC detector reference design report (Ref-TDR) has been released in Oct. 2025. CEPC plans to apply for the starting the construction in the 16th five year plan (2030-2035).

## Introduction

Uncovering the fundamental laws of the universe and its evolution is a great endeavour for humanity. The most effective way to achieve this goal in particle physics is via powerful, high-energy accelerators. In the last half century, many fundamental particles have been found by electron-positron colliders and hadron accelerators, such as charm quark in 1974, bottom quark in 1977, W and Z bosons in 1983, top quark in 1995. And the corresponding “factories” have been built to make detailed studies, such as BEPC/BEPC-II/VEPP-4 for tau-charm factories , LEP/SLC for Z factories, LEP-II for Z and W factories, and KEK B/PEP-II/Super KEK B for bottom factories, etc. [1]

On July 4, 2012, with the discovery of the Higgs boson with a mass of around 125 GeV at the Large Hadron Collider (LHC) at CERN, human being entered the new era of the “Higgs boson” and the door to the unknown part of the universe is wide open. The Higgs boson is the unique “elementary particle” that has zero-spin and participates in multiple non-gauge interactions in the SM. The Higgs boson has a natural and profound connection to the SM “questions” and “defects”, for example the dark matter and the antimatter asymmetry in the universe, and offers one of the best windows to explore the new physics. It is natural, important and urgent to construct Higgs factories somewhere in the world to echo these very fundamental physics research demands. For these future large accelerators, we need new paradigm/theory, new colliders equipped with advanced detectors.

In Sept. 2012, Circular Electron Positron Collider (CEPC) as a Higgs factory constructed in China was proposed by Chinese scientists. As a staging possibility, a Super proton-proton

* Jie Gao: gaoj@ihep.ac.cn

Collider (SppC) was proposed also to be constructed in the same tunnel beside the CEPC, with the later ep and eA collision possibilities to explore the physics experimental potential in an maximum way of CEPC-SppC collider complex [2]. In addition to CEPC, CERN has proposed Future Circular Collider (FCCee and FCChh); Japan Association of High Energy Physicists (JAHEP) proposes to construct a 250 GeV center-of-mass ILC promptly as a Higgs factory with energy upgrade potential to 1TeV; US has proposed C3 Higgs factory and envisaged in future a 10TeV muon collider. The particle physics community worldwide reached a consensus that Higgs factories must be constructed to guarantee the brave exploration to understand the origine of universe and its evolution in general.

Among circular and linear Higgs factories, they are quite in complementary in luminosities and energy reach potentials. The common challenges for these future large collider proposals are physics problem researches, advanced accelerator and detector technologies, such as SC accelerator technologies, positron source, damping ring, final focus, MDI, cryogenic system, sustainability, industrial promotion and participation, and outreach activities, etc. The collective efforts and collaborations are needed to stimulate each other to strive towards the common goal.

## CEPC general status

The CEPC accelerator complex comprises four accelerators: a 30 GeV Linac, a 1.1 GeV Damping Ring, a Booster capable of achieving energies up to 180 GeV, and a Collider operating at varying energy modes (Z, W, H and top pair). The Linac and the Damping ring are situated on the surface, while the Booster and Collider are housed in a 100 km circumference underground tunnel, strategically accommodating future expansion with provisions for a potential Super proton-proton Collider (SppC), as shown in Fig. 1. The CEPC primarily serves as a Higgs factory. In its baseline design the synchrotron radiation power of 30 MW per beam, it can achieve a luminosity of $5\times10^{34}$cm$^{-2}$s$^{-1}$ per interactions point (IP), producing an integrated luminosity of 13 ab$^{-1}$ for two IPs over a decade, corresponding to 2.6 million Higgs boson events. If the single beam power is increased to 50 MW the CEPC's capability will expand to produce 4.3 million Higgs events, facilitating expected measurements of Higgs coupling at sub-percent levels, exceeding the precision expected from the HL-LHC by an order of magnitude for many final states of the Higgs.

After the completion of the Conceptual Design of the accelerator and the detectors [3,4] in 2018, the CEPC accelerator Technical Design Report (TDR) [5] was finalized in December, 2023, following comprehensive reviews in technical systems in June, and the cost review in Septembers by two separate groups of leading international experts, respectively. The designs and the costs of civil engineering were presented to a review committee consisting of domestic experts, the outcome of which was reported to an international panel which in turn briefed the international cost review committee and the CEPC international advisory committee (IAC). The TDR has been approved by all the review committees and endorsed by the IAC. This TDR details the machine's layout, performance metrics, physical design and analysis, technical systems design, R&D and prototyping efforts, and associated civil engineering aspects. Also the cost estimates (36.4B RMB, ~5B USD) and a preliminary construction timeline are included.

Since Jan. 2024, CEPC entered into Engineering Design Report (EDR) phase (2024-2027), with the key milestones of completion of industrial fabrication of accelerator components, civil engineering design, CEPC EDR site geological investigation, etc. with the aim of applying to start the construction during the 16$^{\text{th}}$ five year plan (2030-2035).

The CEPC group has been working on a reference technical design report (Ref-TDR) of a detector system for the CEPC experiments. The Ref-TDR is mainly intended for project review and approval in China to complete the overall CEPC complex. Once the project is approved, calls for letter of intents will be initiated to form genuine international collaboration and the formal TDR of the detectors will be worked on by the detector collaborations. The CEPC detector Ref-TDR has been released in Oct, 2025 [6] with international reviews by the International Detector Review Committee (IDRC).

In the last decade, the CEPC R&D programs covering the optimization designs, full spectrum technologies and components in accelerators and detectors, studies on software and physics performance, site selections, geological investigation and civil engineering designs, etc. have been in steady progress in synergy with the construction of a 6GeV 4th generation high performance light source by IHEP at its Huairou campus, which has been completed in Oct. 2025, as well as with various detector upgrade projects, such as HL-LHC , participated by the CEPC collaborators,.

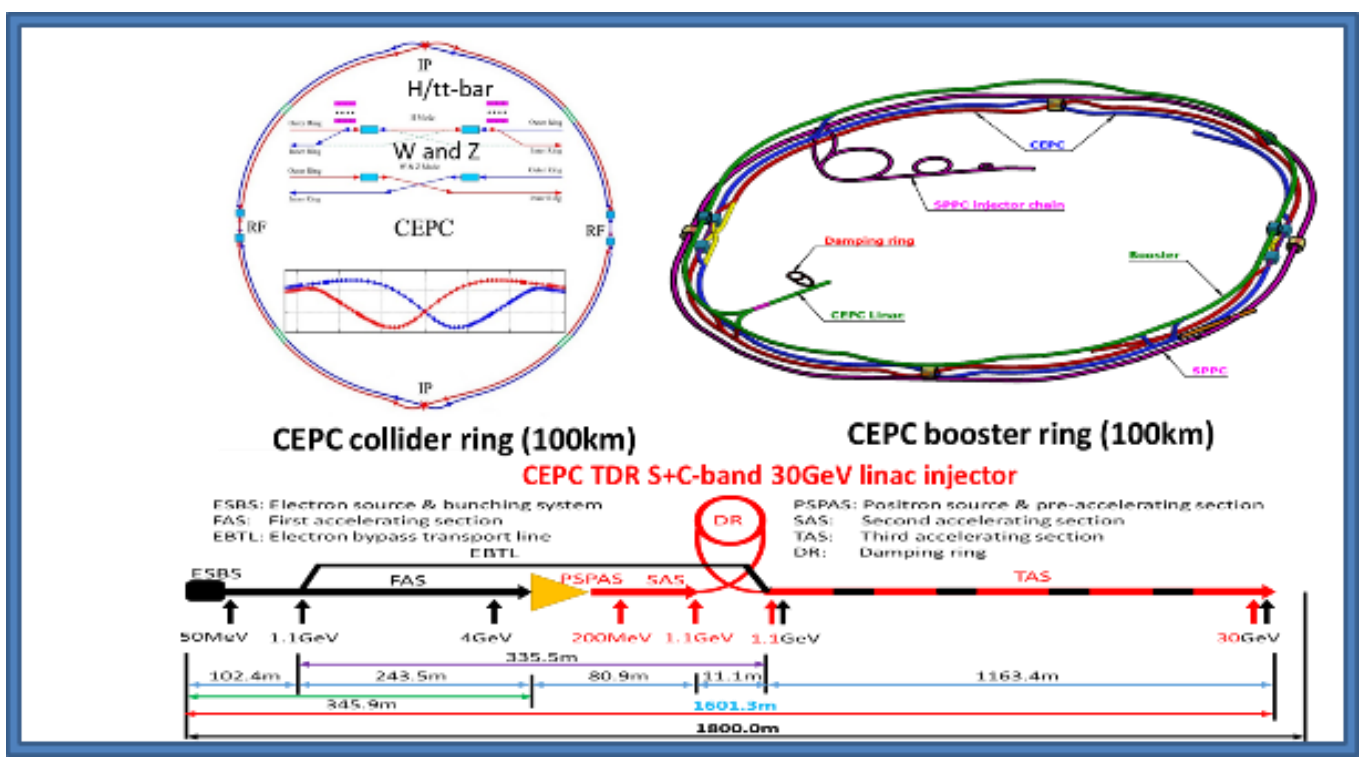

Fig. 1. CEPC-SppC layout.

## CEPC and SppC Luminosities

The CEPC collider design are based on crab-waist collisions, and the maximum luminosities at different energies could also be estimated by J. Gao's analytical formulae[7][8][9] including crab waist effects. The general maximum luminosity formula of colliders per IP of a circular electron positron collider is shown in Eq. (1)[8][9] as follows:

$$L_{max}[\mathrm{cm^{-2}s^{-2}}] = 0.158 \times 10^{34} \frac{(1+r)}{\beta_y[\mathrm{mm}]} I_b[\mathrm{mA}] \sqrt{\frac{U_0[\mathrm{GeV}]}{N_{IP}}}\, e^{\left(\frac{\sqrt{\Phi_p}}{3.22} - \frac{A[mm]}{\beta[mm]_y^2}\right)} (1 + 0.000505\Phi_p^2) \qquad (1)$$

where $r=\sigma_y/\sigma_x$ and $\beta_y$ are the values at interaction point (IP), $U_0$ is the synchrotron radiation energy loss per tune of a charged particle, $N_{IP}$ is the number of IP, $I_b$ is the beam current, $\Phi p$ is the Piwinski angle, $A$ is a machine dependent parameter, and $\frac{A[mm]}{\beta[mm]_y^2}$ is related to the square of the ratio of required injection acceptance to the beam dynamic aperture at the injection position, where for Super KEK B, $A$=1.6mm$^2$, and the $A$ of CEPC will be much smaller due to the on axis injection scheme adopted. For a collider of isomagnetic lattice like CEPC, by using $U_0=C_\gamma E^4/R$ and the beam power $P_b=I_bU_0$, Eq. (1) could be expressed also in Eq. (2) as follows:

$$L_{max}[\mathrm{cm^{-2}s^{-2}}] = 0.158 \times 10^{34} \frac{(1+r)}{\beta_y[\mathrm{mm}]} \sqrt{\frac{R[\mathrm{m}]}{C_\gamma[\mathrm{mGeV^3}]N_{IP}}} \times$$

$$\left(\frac{P_b[\mathrm{MW}]}{E[\mathrm{GeV}]^2}\right) e^{\left(\frac{\sqrt{\Phi_p}}{3.22}-\frac{A[mm]}{\beta[mm]_y^2}\right)} (1 + 0.000505\Phi_p^2). \quad (2)$$

where $C_\gamma$=8.85×$10^{-5}$mGeV$^{-3}$, R is the local bending radius of dipoles of collider rings and E is the beam energy. For a collider of non-isomagnetic nature, like Super KEK B, Eq. (1) could be used to estimate the maximum achievable luminosity.

As for a hadron collider, such as SppC, $FCC_{hh}$ or LHC, the maximum beam-beam tune shift value per IP could also be estimated by J. Gao's analytical formulae[7][8], both for round and flat beams, as expressed in Eq. (3)[8]:

$$\xi_{max,y} = \frac{H_0\gamma}{f(x_*)}\left(\frac{r_p}{6\pi R N_{IP}}\right)^{1/2} F \quad (3)$$

where $H_0$=2845, $\gamma$ is the normalized beam energy, $r_p$ is the hadron (proton) radius, R is the local dipole bending radius, $N_{IP}$ is the number of interaction point, F= (4√2)/3 for round beam collision, F=1 for flat beam collision, f(x) is expressed in Eq. 4, and $x_*$ can be solved by Eq. (5).

$$f(x) = 1 - \frac{2}{\sqrt{2\pi}}\int_0^x e^{-\frac{t^2}{2}}dt \quad (4)$$

$$x_*^2 = \frac{4f(x_*)^2}{H_0\pi\gamma}\left(\frac{6\pi R}{r_p N_{IP}}\right)^{1/2} \quad (5)$$

By using the parameters of SppC[5] and Eqs. (3)-(5) for round beam collision, the analytically estimated $\xi_{max,y}$ is 0.0147, and the numerical estimation value of $\xi_{max,y}$ for SppC is 0.015. As for LHC, the analytical estimated $\xi_{max,y}$ value is 0.0045 by using Eqs. (3)-(5)[7][8], and the LHC parameter table $\xi_{max,y}$ value is 0.005, and the experimental result is about 0.0045[10].

## CEPC Accelerator Industrial Preparation in EDR

In CEPC accelerator EDR phase, industrial fabrication preparations of some key components are under developments, and some examples will be given in the following. The booster dipole magnets' automatic fabrication line, which could fabricate 4 magnets of 5 meter long per day to guarantee the construction time and decrease fabrication cost, is shown in Fig. 2, and it will be completed in 2025.

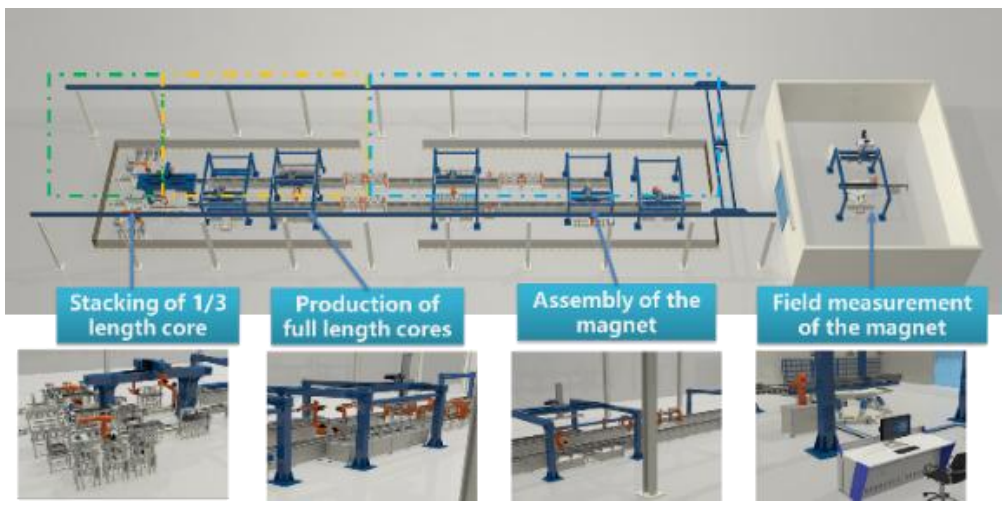

Fig. 2. CEPC booster magnet automatic fabrication line.

CEPC collider rings have about 200km coated copper vacuum chambers, and the NEG coating vacuum chamber automatic production line will greatly increase NEG coating efficiency and reduce the fabrication cost as shown in Fig. 3, which will be completed in 2025. This demonstration automatic production line could produce 2 NEG coated vacuum chambers of 11.5 m long, and in final production for CEPC construction, 9 lines will be needed.

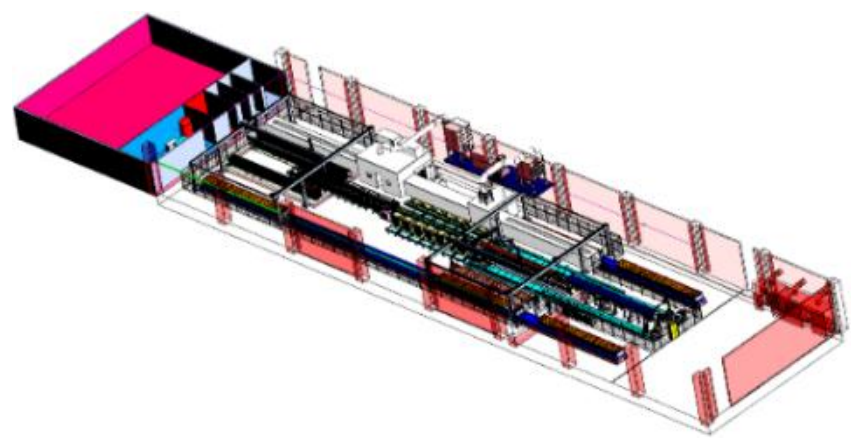

Fig. 3. CEPC collider ring NEG coating vacuum chamber automatic production line.

For CEPC collider ring, a 650 MHz full-scale cryomodule prototype is under fabrication to achieve CEPC EDR specification as shown in Fig. 4. The cryomodule will contain six 2-cell cavities with the operating gradient of 25 MV/m and quality factor of $3\times10^{10}$. The cavity will adopt state-of-the-art medium temperature baking high Q high gradient technologies. The input couplers will be variable to match the different operation conditions and have more than 300 kW CW power capacity. HOM couplers and two 5kW absorbers will be installed as well. Operating at 2K, the static heat load of the cryomodule is designed to be less than 30W. The cryomodule horizontal test will be completed by June 2026 on PAPS of IHEP.

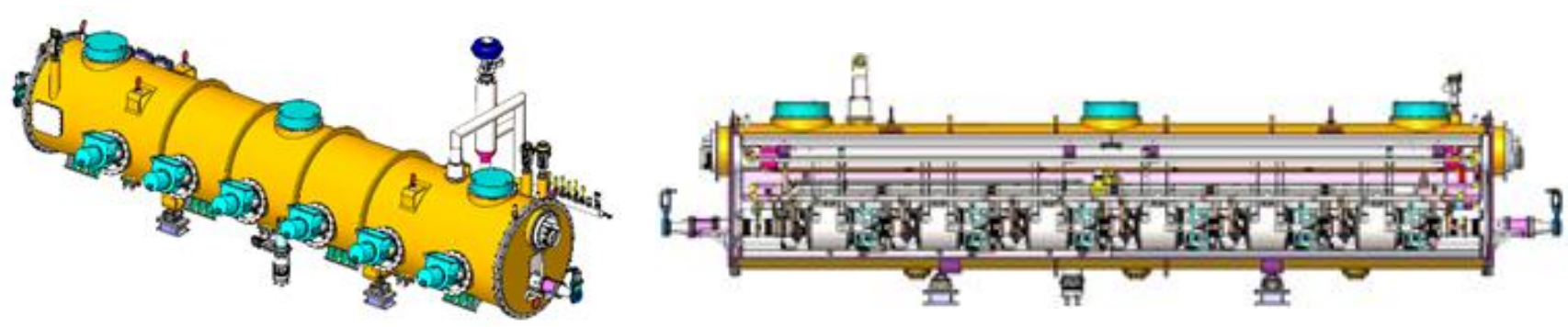

Fig. 4 CEPC collider ring 650MHz full-scale cryomodule.

In CEPC EDR phase, different types of klystrons will be developed continuously. In 2024, a 650MHz CW 800kW high efficiency klystron has reached the efficiency of 78.5%, as shown in Fig. 5 (left). A 650MHz multibeam klystron of designed efficiency of 80.3% will be tested in 2025 as shown in Fig. 5 (right) under degassing process. To continue increasing the 650MHz klystron's efficiency, a new type of energy recovery klystron will be developed with one energy recovery stage, the efficiency could reach 85% as shown in Fig. 6 (left). If two stages are used the efficiency could reach 90%. Concerning the C-band klystron for CEPC injection linac, a klystron of 5720 MHz, 80MW peak power, 3us pulse length, and 100Hz repetition rate, is under construction, as shown in Fig. 5 (right), which will be tested in 2025. By using this 80MW C-band klystron, a C-band test band will be developed to test various components of CEPC C-band linac, such as high gradient C-band accelerating structure of 40~45M/m.

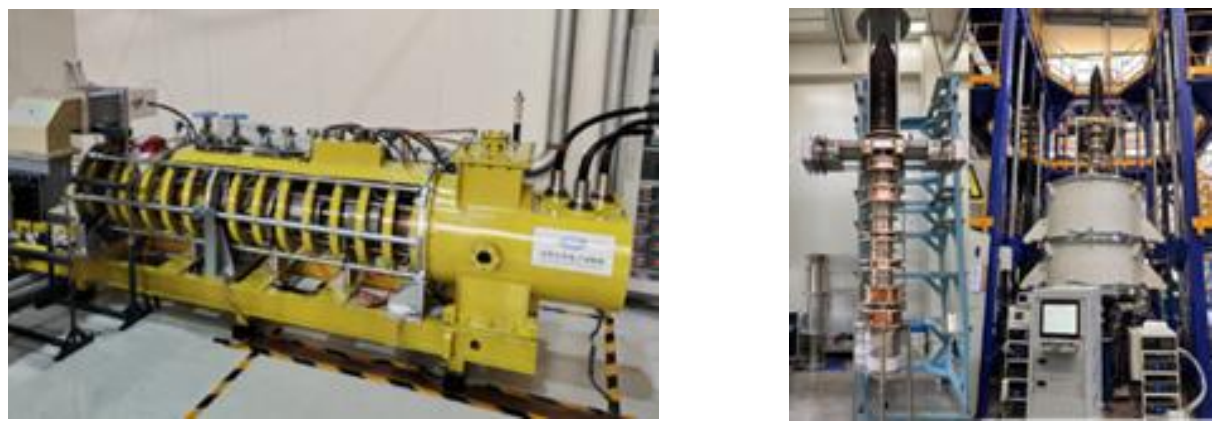

Fig. 5. CEPC 650MHz 803kW CW klystron with efficiency 78.5% achieved in 2024 (left); CEPC multibeam 650MHz 800kW CW klystron with efficiency 80.3% (right).

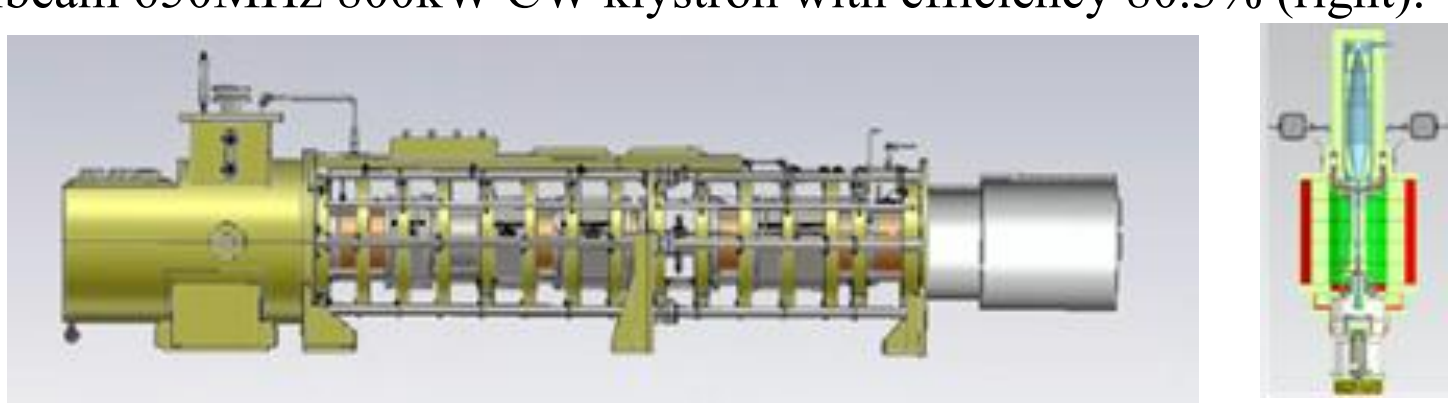

Fig. 6. CEPC 650MHz one stage energy recovery klystron with efficiency of 85% (left); CEPC C-band klystron of 5720 MHz, 80MW, 3us pulse length and 100Hz repetition rate (right).

Concerning CEPC detector TDR reference design, many progresses have been made in pixel vertex detector, tracker & PID, calorimeters, muon detector, high field detector SC magnet, lumical, MDI, software, TDAQ, detector mechanics, etc., as shown in Fig. 6. Some of the sub-detectors, such as vertex detector and calorimeters, have been beam tested at DESY and CERN test beam stations in 2022, 2023 and 2024, respectively. As an important milestone, the CEPC detector TDR reference design report has been released in Oct. 2025[6].

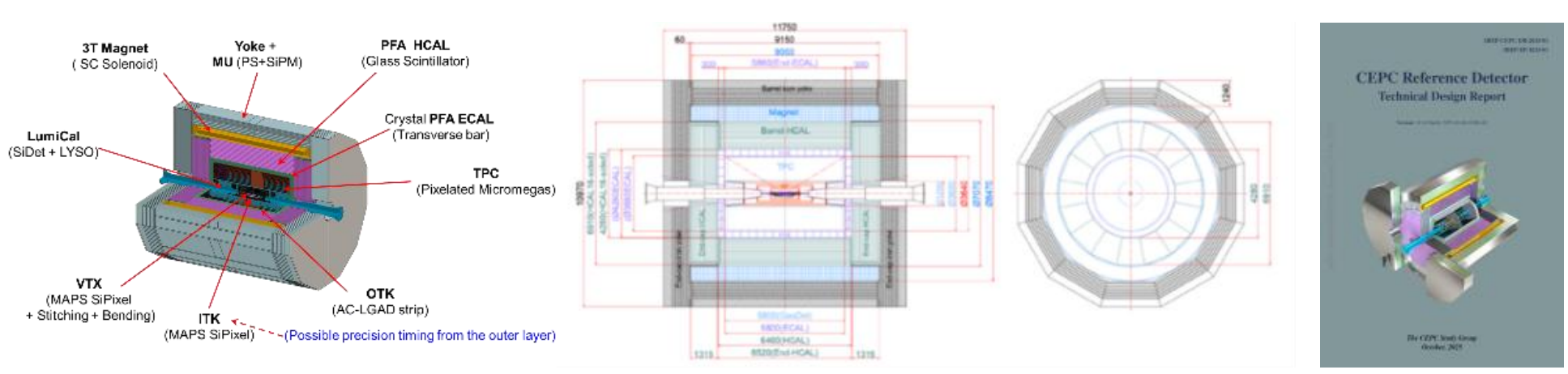

Fig. 8. CEPC TDR detector reference design layout.

## CEPC alternative technologies development

Plasma acceleration is a very promising strategic technology for future high energy physics thanks to its accelerating gradient as high as more than three order of magnitude compared with that of the conventional radio frequency accelerator technology. Since last 40 years, plasma acceleration has made enormous progresses, between laser driven and beam driven plasma acceleration technologies, the later one has important advantage of much high efficiency from AC power to the accelerated beam which is vital for high energy colliders. Apart from the beam driven plasma acceleration, the staging and positron acceleration are other key technologies need to be demonstrated with the required high beam qualities in practical applications. In 2023, IHEP started a beam driven plasma acceleration experiment test facility development with the financial support from CAS by using the BEPC-II electron and positron injection linac of 2GeV. To demonstrate staging and positron acceleration technologies, a new beam line of high bunch charge equipped with an L-band photo cathode RF gun parallel to BEPCII linac will be constructed. Positron beam will be accelerated by using hollow channel plasma. The intermediate application of such a beam driven plasma accelerator is a CEPC plasma injector for booster[11], and for the future applications, plasma based electron positron linear colliders are the main goals .

Polarization beam technology in high energy colliders is very challenging, both for linear collider and circular collider, but especially for circular one. In circular colliders, such as CEPC, 5%~10% transversely polarized beam can be used to precisely measure the beam energy, and the ~50% longitudinal colliding beam could be vital for some specific experimental channels. For weak transverse polarized beams, self-polarization scheme can be used in the collider ring by using special unsymmetrical wigglers without polarization injection from the booster. For longitudinal collision polarized beams, one needs polarized beam source from the linac injector. Due to the large circumference of CEPC booster, it is found that during the energy ramping, it is not necessary to use Siberian snakes to avoid the depolarization resonances, and preliminary studies show that for Z and W energies, both the transverse and longitudinal polarization beams could be achieved, and the Higgs energy case is still under study.

## The SppC

The SppC presented in the TDR consists of two IPs with center-of-mass energy reaching 125 TeV in pp collision with 20T high field dipole magnets with a luminosity target of $10^{35}cm^{-2}s^{-1}$ per IP. The main progresses for SppC are with the long term and persistent development of High Field Magnet (HFM) particularly in the areas of the Iron Based High Temperature Superconducting (HTS) material and the magnet dipole, and the design consistency of the CEPC Higgs factory with the future requirement of the SppC. The group and its collaborators have achieved a highest quenched field of 14T@4.2K with the model dipole magnet with the Nb3Sn+HTS combinations in 2023. A 16T@4.2K model dipole is under development. According to the study progresses of the Iron Based Superconducting (IBS) material performances in the last 8 years, it is expected that IBS material could reach the application goal of >20T around 2035.

## CEPC international and industrial collaborations

CEPC has always been envisioned as an international big-science project and the international colleagues played a significant role in the study of CEPC CDR and TDR. A total of 1143 authors from 221 institutes (including 144 foreign institutes) across 24 countries and regions co-signed the CEPC CDR and a total of 1114 authors from 278 institutes (including 159 foreign institutes) across 38 countries and regions co-signed the CEPC accelerator TDR. As for the CEPC detector reference TDR, it has been released with >1470 authors, >30% international authors, >384 institutes, >63% international institutes, and >43 countries. The CEPC study group has signed more than 21 MoUs with institutes and universities from Europe, Asia, the Americas, Africa, and Oceania. The CEPC team organizes CEPC international workshops and European/US versions with significant international participations.

The CEPC's development has been operated in an international way and guided by international committees, such as International Advisory Committee (IAC), the International Accelerator and Detector Review Committees.

Concerning industrial collaboration, since November 2017, CEPC Industrial Promotion Consortium (CIPC) has been established and more than 100 companies have joined, and CIPC has played an important role in the CEPC TDR and EDR processes both for accelerator and detector.

## Summary

The CEPC is on the path to converge towards construction. The present EDR process will bridge the TDR technologies (both accelerator and detectors) to industrialization for final construction and bring benefits to the HEP community in general. CEPC will apply to Chinese government to start the construction during the $16^{th}$ five year plan (2030-2035).
As an international project, the CEPC has received a great deal of participations and helps from international collaborating scientists and institutions, and the project is committed to maximize international collaboration and contribution. International efforts towards Higgs factories are common endeavours of human beings, and the final goals should be and could be reached with persistence and endurance. The CEPC Study Group is making strong efforts towards offering the worldwide HEP community an early Higgs factory.